\documentclass[11pt]{cas-sc}
\usepackage[utf8]{inputenc}
\usepackage{setspace}
\usepackage{amsmath,amssymb,amsfonts}
\usepackage[authoryear]{natbib}
\usepackage{parskip}
\usepackage{graphicx}
\usepackage{array}
\graphicspath{ {./figures/} }
\usepackage{caption}
\usepackage{subcaption}

\usepackage{xcolor}
\usepackage[colorinlistoftodos,prependcaption,textsize=tiny]{todonotes}

\usepackage[mathlines]{lineno}
\begin{document}

\title[mode = title]{
%Deep-Learning Emulator: Paper1 (suggested title: 
A novel deep learning approach for emulating computationally expensive postfire debris flows}

\shorttitle{A novel deep learning approach for emulating computationally expensive postfire debris flows} 

\author[1]{Palak Patel}[type=editor, 
                        orcid=0009-0003-9786-1359]%\email{}
\shortauthors{P. Patel et al.}

\author[2]{Luke McGuire} %\email{} 

\author[3,1]{Abani Patra}
\cormark[1]
\ead{abani.patra@tufts.edu}

\cortext[cor1]{Corresponding Author}

\address[1]{Department of Mechanical Engineering, Tufts University, Medford, MA-02155}
\address[2]{Department of Geosciences, University of Arizona, Tucson, AZ-85721}
\address[3]{Department of Computer Science, Tufts University, Medford, MA-02155}

\begin{abstract}
   % Geophysical flows, such as debris flows and landslides, pose significant risks to human lives and infrastructure, necessitating accurate and computationally efficient simulations. 
    Traditional physics-based models of geophysical flows, such as debris flows and landslides that pose significant risks to human lives and infrastructure are computationally expensive, limiting their utility for large-scale parameter sweeps, uncertainty quantification, inversions or real-time applications. This study presents an efficient alternative, a deep learning-based surrogate model built using a modified U-Net architecture to predict the dynamics of runoff-generated debris flows across diverse terrain based on data from physics based simulations. The study area is divided into smaller patches for localized predictions using a patch-predict-stitch methodology (complemented by  limited global data to accelerate training). The patches are then combined to reconstruct spatially continuous flow maps, ensuring scalability for large domains. To enable fast training using limited expensive simulations the deep learning model was trained on data from an ensemble of physics based simulations using parameters generated via Latin Hypercube Sampling and validated on unseen parameter sets and terrain, achieving maximum pointwise errors below 10\% and robust generalization. Uncertainty quantification using Monte Carlo methods are enabled using the validated surrogate, which can facilitate probabilistic hazard assessments. This study highlights the potential of deep learning surrogates as powerful tools for geophysical flow analysis, enabling computationally efficient and reliable probabilistic hazard map predictions. 
\end{abstract}


\maketitle

% For Review and Manuscript Submission
% \doublespacing 

\section{Introduction}
Computational models that can efficiently simulate geophysical flows such as volcanic flows, avalanches, landslides, and debris flows, are crucial for hazard and risk assessments. Outputs from these models, including area inundated and peak flow depth, can be used to identify flow runout paths, predict building damage, and evaluate potential hazard mitigation or risk reduction strategies \citep{vallis2016geophysical,barnhart2024evaluation,mccoy2016minimizing}.
These simulations can account for flow parameters, such as flow rheology, volume, and initiation location, as well as topography and surface characterization to quantitatively assess flow hazards. Information regarding the probability of flow hazards can further enhance early warning capabilities and evaluation of mitigation strategies \citep{oakley2023toward,barnhart2023user}, especially where the historical record of past flow events is inadequate.

Titan2D, a geophysical flow modeling software,  simulates flow over topography using a depth-averaged shallow flow model with complex rheologies \citep{titan2d2005}
%. %Titan2D uses advanced numerical methods to efficiently solve hyperbolic systems of equations and is designed for parallel computing on multiple processors, which decreases computing time for flow simulations. Titan2D simulations 
providing descriptions of flow propagation across a landscape given inputs that define initiation conditions, topography, and flow rheological parameters. Despite efforts to improve computational efficiency, Titan2D and similar geophysical flow models, are computationally expensive. Run times for simulations can range from minutes to a day, depending on the size of the domain map, level of detail (e.g., desired spatial resolution), computing resources and complexity of the physics modeled in the simulation. %Additionally, a large number of input parameters can be needed to define flow physical and rheological properties. 
This makes it infeasible to use such simulations for uncertainty quantification(UQ) or probabilistic hazard assessments that rely on Monte Carlo techniques requiring thousands to millions of simulations \citep{Daw_Karpatne_Watkins_Read_Kumar_2021}.

Cost-effective surrogate models are thus a required element of  simulation-based hazard analysis for this problem which typically requires the evaluation of  large model ensembles corresponding to different choices of inputs and parameters for the Titan2D code\citep{Bayarr_Berger_Calder_Dalbey_Lunagomez_Patra_Pitman_Spiller_Wolpert_2009, diniz2025efficient, emerick2013ensemble}. Such models are also invaluable for model sensitivity, data assimilation/calibration and uncertainty quantification.
%A surrogate model is a simplified representation of a complex system that can be used to approximate the output of the system without needing to run expensive or time-consuming simulations. 
Surrogate models enable rapid explorations of flow inundation area or peak flow depth associated with various parameter settings, as well as for UQ %Surrogate models therefore allow for a more efficient way to analyze and explore the effects of various sources and types of uncertainty, without the limitations of traditional geophysical flow simulations 
\citep{Bayarri2015}.

It is often challenging to use traditional surrogate methods such as those based on Gaussian Processes to
%or polynomial regression to 
 capture complicated relationships in high dimensional parameter spaces when modeling spatiotemporal flow structures with complex physics and localized structures \cite{Bayarri2015}. %\citep{sun2020surrogate}. 
This is largely due to the difficulty of finding suitable kernels and covariance structures that represent dependencies and the rapid increase in the computational cost of these surrogate modeling techniques with large data ($\cal{O}$$ (n^3)$ cost of the inverses of covariance matrices) necessary for capturing fine scale flow structures\citep{sun2020surrogate}. Deep learning methods provide an interesting alternative as surrogate models, especially in areas where traditional surrogate models have limitations. Availability of adequate data (from simulation outcomes in this case) allows us to build satisfactory surrogates without explicit representations. These techniques are effective at approximating complex relationships in various applications, including image and speech recognition, natural language processing, and computer vision \citep{doi:10.1142/11389-vol1}. Deep learning approaches can also be particularly useful in simulating complex mechanistic processes 
%due to their ability to learn and represent non-linear relationships between inputs and outputs, which is often required for modeling the underlying physics of these processes 
in fluid dynamics and earth sciences \citep{tripathy2018deep, Reichstein_Camps-Valls_Stevens_Jung_Denzler_Carvalhais_Prabhat_2019}. Deep learning, especially when employing convolutional neural networks, has been shown to be useful for flow hazard prediction due since it can extract significant spatiotemporal features from data \citep{Löwe_Böhm_Jensen_Leandro_Rasmussen_2021}.

Surrogate modeling based on deep learning for complex geophysical flows still faces challenges largely due to  
 the high dimensionality of the input parameter space, the presence of uncertainties in the input parameters, and the limited availability of high-resolution input data (e.g., topography) can make it difficult to train accurate deep learning models. 
%Second, deep learning models trained on a specific dataset may not be transferable and can struggle to make accurate predictions when applied to locations not included in the training datasets (\citep{Stolp_2021, Berkhahn_Fuchs_Neuweiler_2019}).
%Third, deep learning models can be architecture-specific in terms of input and output size, with a need to retrain the model when changing the size of the simulation domain. One solution is to train a separate deep learning model for each location and domain of interest. However, this solution poses a serious limitation since it would require large training data for each area of interest, which is computationally expensive and time-intensive (\cite{chen2019unet}). 
We mitigate these challenges through our choice of architecture and training strategy.

In this study, we develop a deep learning-based surrogate modeling framework to emulate geophysical flows using a modified U-Net architecture with skip connections, a deep learning network that is widely used for image segmentation and has been shown to be effective for computational fluid dynamics \cite{ao_li24,chen2019unet}. The framework uses Titan2D simulations to create training data to estimate the peak flow depth along the runout path of a debris flow. 
%This circumvents the need to obtain training data from field-based observations, which is expensive, logistically challenging, and time consuming. 
Limited field data are used to calibrate the simulations (\cite{Spiller_McGuire_Patel_Patra_Pitman_2024}) while the simulator is used to generate ``training data". The proposed modified U-Net architecture can be trained on a limited number of simulations often numbering in the tens when they are cleverly augmented. Thus, the primary objective of this study is to develop a deep learning based network model that can rapidly and accurately predict 2D maps of maximum flow depth based on training data from a limited number of physics-based simulations,  enabling UQ  and probabilistic hazard analysis as in \cite{Bayarr_Berger_Calder_Dalbey_Lunagomez_Patra_Pitman_Spiller_Wolpert_2009}.

\section{Methods}

\subsection{Study Areas}

\begin{figure}[h!]
    \centering
    \includegraphics[width=\textwidth]{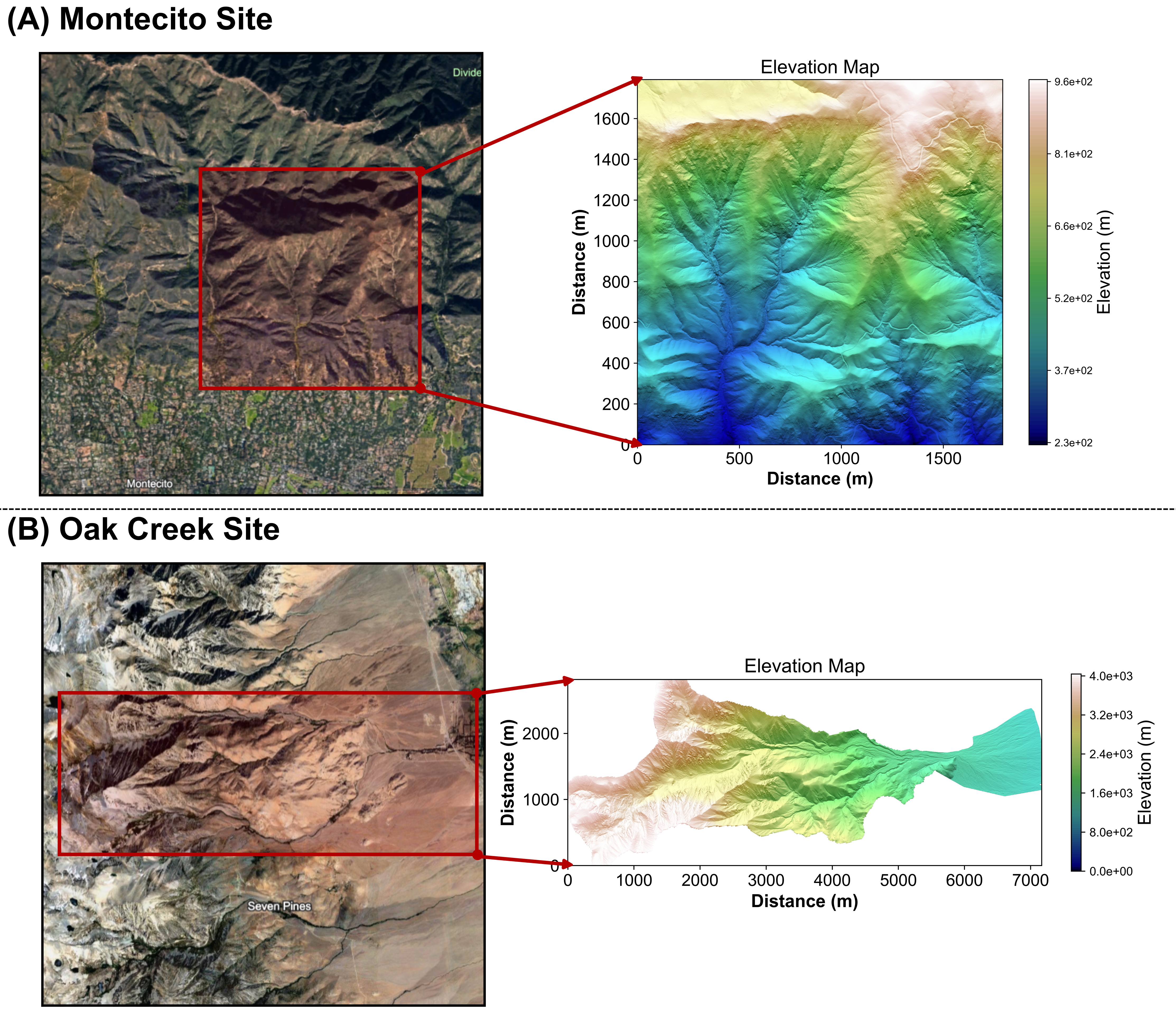}
    \caption{Study Areas for Model training and evaluation. (A) The study area near Montecito, California, USA. (B) The study area encompassing the Middle Fork of Oak Creek, near Independence, California, USA. The satellite image on the left shows the study regions. The red box highlights the region with the corresponding DEMs used for deep learning model evaluation on the right. The satellite images are sourced from Google Maps \citep{googlemaps_montecito, googlemaps_oakcreek}.}
    \label{fig:study_area}
\end{figure}

We simulate debris flow initiation and runout at two topographically dissimilar  study areas. We first develop and test our proposed emulator at a site in southern California, USA, that we refer to as the Montecito study site (Figure \ref{fig:study_area}). The second study area, which we refer to as Oak Creek is located in the Eastern Sierra Nevada near Independence, California, USA. We apply our emulator methodology at this site as a case study to illustrate its broader applicability regardless of the specific terrain at any one site.

Both the Montecito and Oak Creek sites were burned by wildfires and then produced runoff-generated debris flows during post-fire rainstorms \citep{kean2019inundation,degraff2011remarkable, wagner2012oak}. Runoff-generated debris flows initiate when overland flow rapidly entrains sediment from hillslopes and steep channels \citep{kean2013runoff,wells1987effects}. Steep, burned landscapes are more susceptible to runoff-generated debris flows relative to unburned areas since fire  temporarily enhance runoff and erosion during rainstorms by reducing soil infiltration capacity and decreasing ground cover \citep{robichaud2016infiltration,larsen2009causes,mcguire2024fire}. We do not attempt to reproduce specific characteristics (e.g., area inundated or peak flow depth) of the observed debris flows at either of these sites. Instead, we use these sites for model development, testing, and application since they are broadly representative of steep landscapes that are susceptible to postfire runoff-generated debris flows in the western USA, a region where fire activity is increasing \citep{singleton2019increasing}. The similarities in the debris-flow generation processes at both sites, namely that debris flows initiated from runoff that rapidly entrained sediment, also allows us to use the same set of governing equations for physics-based simulations at both sites.

The Montecito site experienced debris flows on January 9, 2018, when an intense rainstorm impacted the Santa Ynez Mountains. \citep{kean2019inundation,alessio2021post,morell2021sediment}. Debris flows initiated in watersheds located above the community of Montecito, California, that had recently burned in the Thomas Fire. The debris flows traveled across an alluvial fan for several kilometers, leading to loss of life and substantial damage \citep{lancaster2021observations}. Several of the most impactful debris flows affected the Montecito, San Ysidro, Buena Vista, and Romero Creek watersheds \citep{kean2019inundation}.
In this study, we focused on modeling debris flows within upper San Ysidro Creek and a smaller, adjacent watershed (Figure \ref{fig:study_area}). Specifically, our area of interest includes the steep headwater region above the alluvial fan, where debris flows initiated and grew but remained relatively confined to channels.

A runoff-generated debris flow initiated at the Oak Creek site on July 18, 2008, roughly one year following the 2007 Inyo Complex Fire \citep{degraff2011remarkable}. The debris flow traveled 6-7 kilometers downstream from the mountain front, damaging 50 residential structures \citep{wagner2012oak}. Our model domain consists of the Middle Fork of Oak Creek, which was a major contributor of sediment for the observed debris flow \citep{wagner2012oak}, and a portion of the downstream alluvial fan. Therefore, in contrast to the Montecito site, the model domain includes the steep headwater region where flows were relatively confined as well as part of the more gently-sloping alluvial fan surface where flow spreading and bifurcation were more common (Figure \ref{fig:study_area}).

\subsection{Simulating Debris Flow with Titan2D}
%\improvement{Do we need to add an appendix on the equations? }
The initiation and runout of runoff-generated debris flows can be simulated using morphodynamic models that represent infiltration, runoff, sediment transport, and changes to flow resistance associated with spatial and temporal variations in sediment concentration \citep{McGuire_Kean_Staley_Rengers_Wasklewicz_2016,mcguire2017debris,Spiller_McGuire_Patel_Patra_Pitman_2024}. Simulations are typically run at the event time scale, such as an individual rainstorm, and can be used to explore the relationships between landscape properties, rainfall characteristics, and the resulting debris flow initiation, growth, and runout processes \citep{mcguire2021time,Spiller_McGuire_Patel_Patra_Pitman_2024}. Simulation results are dependent on a relatively large set of parameters that influence rainfall-runoff partitioning, rates of sediment detachment by runoff and raindrops, sediment deposition rates, and flow resistance. The model has been applied to simulate postfire sediment transport and debris-flow initiation in small ($< 1$ $\mathrm{km}^2$) watersheds in southern California, USA \citep{McGuire_Kean_Staley_Rengers_Wasklewicz_2016,mcguire2017debris,tang2019evolution,mcguire2021time}. This model was recently integrated with Titan2D to facilitate more efficient simulation of debris flows on larger spatial scales \citep{Spiller_McGuire_Patel_Patra_Pitman_2024}. The increase in computational efficiency is due to Titan2D's architecture, including the capability for adaptive mesh refinement and parallel computing \citep{titan2d2005, titan2d2019}. The model, as integrated into Titan2D, has been applied to simulate the entirety of the postfire debris-flow life cycle from runoff generation to debris flow growth and runout at the Montecito site, where it can reproduce the general spatial pattern of inundation on the alluvial fan  \citep{Spiller_McGuire_Patel_Patra_Pitman_2024}. The use of Titan2D enabled us to efficiently generate maps of peak debris flow depth, which we use to train and test our deep-learning approach. 
%\unsure{Do we need to cite DCLAW and other flow solvers? In principle all our work here will seamlessly work with DCLAW simulations}

While we have used Titan2D as our physical simulation tool, the overall development should be invariant for other computational simulation models like DCLAW \citep{Berger2010TheGS, dclaw_software} that are based on similar physical principles, though the input variables and parameter ranges will need to be selected based on the specifics of the model.

\subsection{Deep Learning Framework}

Deep learning has emerged as a powerful approach for regression problems in Earth Sciences, particularly in the extraction of spatiotemporal features from data. Predicting spatially variable peak flow depths across a landscape can be viewed as a problem of translating images using neural networks, where both the input and output consist of images of terrain data and maximum flow depth, respectively. Convolutional neural networks (CNNs) and their derivatives initially developed for computer vision tasks, have been successfully applied in this domain \citep{chen2019unet, Thuerey_2020}. 

In this study, we develop a deep-learning model that can effectively predict maximum flow depth throughout a landscape for a desired parameter set. This model would train and predict on patches, which are defined as spatially contiguous subsets of the input terrain raster, instead of the entire input terrain raster. These predicted patches from our model can then be stitched back together to form a spatially continuous map of peak flow depth throughout the entire model domain. 

There are several important benefits of building a model that works on patches. First, predicting peak flow depths for the entire terrain map in a single pass through the neural network can be computationally expensive, particularly for large areas. Breaking down the prediction task into smaller patches reduces the size of the network, thereby reducing computational costs. In addition, the patchwise prediction approach enables the model to handle terrain maps of varying sizes and resolutions. As each patch is processed independently, the model can scale to accommodate larger maps without necessitating substantial architectural modifications. Lastly, dividing the simulation results into smaller patches increases the size of the training data. This enhances the model's ability to learn and generalize among similar patches.

\subsubsection{Network Architecture}
The U-Net architecture, %which was first developed for biomedical image segmentation, 
can provide accurate results with limited training data \citep{ronneberger2015u}. It combines an encoder-decoder structure with skip connections to preserve detailed spatial patterns in the output  \citep{ronneberger2015u}. %U-Net architecture is designed to follow an encoder-decoder architecture, where 
The encoder part captures the context and features from the input layer, and the decoder generates the output with precise localization. This architecture is well-suited for predicting fluid flow over terrain where capturing both global context (e.g., broad spatial patterns in flow) and fine detail, including variations in flow depth within a channel and immediately outside the channel banks, can be crucial. The skip connections between the encoder and decoder layers help to preserve spatial information and gradients during the upsampling process of the decoder layer. These attributes of U-Net where it can effectively handle both local and global information, enable it to extract flow features at various scales and share the distinguished features found by the encoder with the decoder field for accurate flow predictions
%. This makes it an attractive architecture for our application 
\citep{chen2019unet,lino2020simulating, Lino_Fotiadis_Bharath_Cantwell_2023}. 
%
%U-Net's ability to learn while conserving intricate patterns makes it  critical to utilize in our work, as the flow field simulated using Titan2D can be sensitive to both large- and small-scale terrain features. 
Here we developed a modified variation of the U-Net architecture that also takes an array of parameter sets as input along with the input terrain raster and predicts a map of maximum flow depth for a given parameter set. 

\begin{figure}[ht!]
    \centering
    \includegraphics[width=\textwidth]{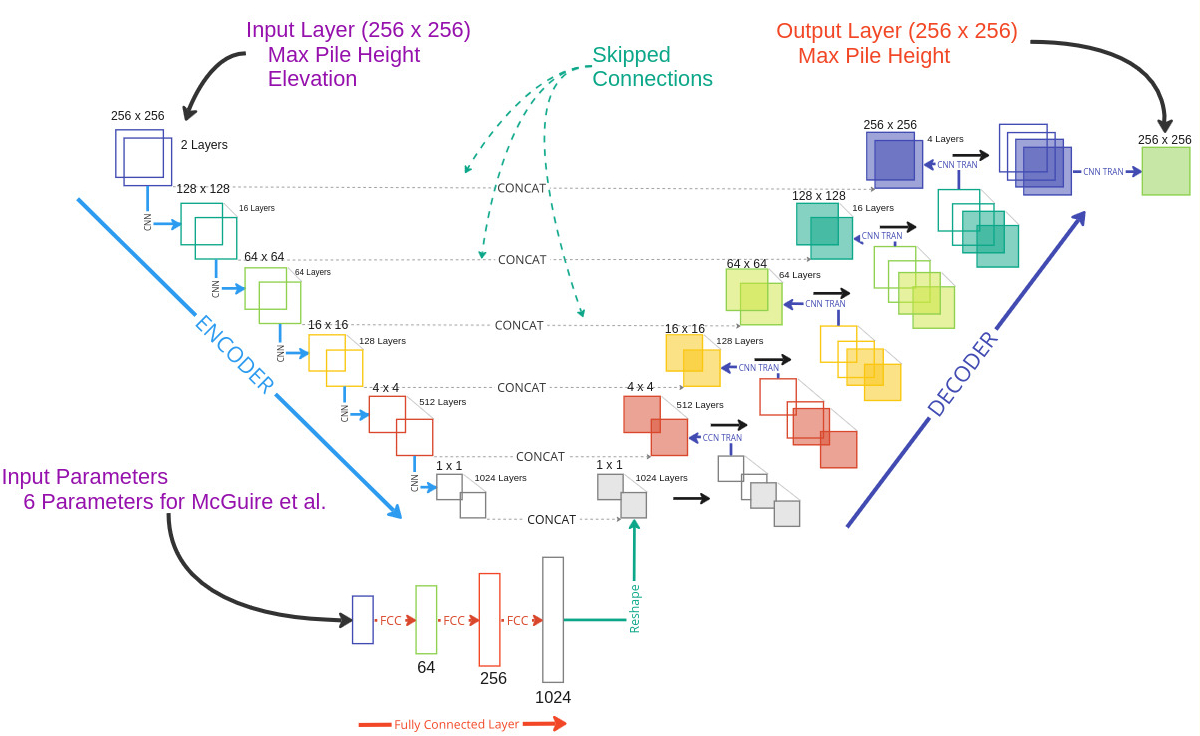}
    \caption{Schematic representation of the modified U-Net architecture, which integrates dual input sets: spatial features (e.g., elevation and max flow height) through the encoder-decoder framework and physical model parameters via fully connected layers. Skip connections ensure spatial feature preservation for accurate predictions of maximum flow depth in the output.}
    \label{fig:net_arch}
\end{figure}

Our modified U-Net network architecture is illustrated in Fig. \ref{fig:net_arch}. Input to our model consists of two separate sets of data. The first set of input is a patch of the terrain raster of size $256 \times 256$ pixels with two channels, terrain data (i.e., a topographic map) and maximum flow depth from a benchmark simulation. The flow depth data from the benchmark simulation is described in the following section. These data are then fed to the encoder layer. 

The encoder layer consists of a series of encoder blocks, each comprising two convolutional layers followed by max-pooling layers to compress the input raster. Similarly, the decoder layer consists of a series of up-sampling blocks using convolutional transpose layers to decompress the compressed arrays and generate the output raster. 

The output array of each decoder block is concatenated with the corresponding output produced by the encoder block, and these concatenated arrays are fed to the successive up-sampling layers. This process of concatenating is called \textit{skip connections}. The skip connections effectively double the number of channels for each decoding block. These skip connections help the network to consider low-level input information learned from the encoder during the reconstruction of the solution in the decoding layers \citep{Thuerey_2020}.

In addition to the terrain raster input, the model takes into account six parameters that can have a substantial influence on simulated peak flow depth and runout extent. To incorporate this into the model, we feed these 6 parameters as input to a fully connected network with three hidden layers which transforms the parameter array of length 6, into an array of length 1024. 
% This resulting array is reshaped into a 3D array of $1 \times 1$ pixels with 1024 channels. 
This array is then reshaped and concatenated with the output of the encoder at the bottleneck of the encoder-decoder network and serves as input to the decoder layer. By combining the parameter information with the encoded features, we improve the model's ability to capture the dependencies between terrain characteristics and parameter settings during the decoding process.

Here we have used a rectified linear unit (ReLU) activation function to introduce non-linearity and enhance the network's representational capacity by enabling the model to capture non-linear relationships \citep{Nair_Hinton_2010}. It is applied after each layer of convolution, convolutional transpose, and dense layer, except for the final output layer of the model. 

\subsubsection{Benchmark Simulation}
A first step in our neural network architecture is to tile the map into small patches. The aim of our model is to take terrain data for each patch as input and predict peak flow depth for the corresponding patch. However, this localized approach inherently lacks global information that may be needed for prediction.
%However, this localized approach inherently lacks direct access to upstream conditions, thereby limiting the model's grasp of the broader context of the terrain and potentially compromising prediction accuracy, particularly in regions where upstream conditions exert significant influence on downstream outcomes.
%
To address this limitation, we present a novel approach. We  include an extra channel in our input layer that contains information from a benchmark simulation. The benchmark simulation was generated with the input parameters set at the midpoint of the range specified in Table \ref{tab:parameter-range}. By incorporating the results of peak flow depth from the benchmark simulation as a reference point, the model not only learns localized features extracted within each patch but also gains a deeper understanding of the broader flow dynamics of the terrain.   
%
%The inclusion of the benchmark simulation increases the model's understanding of the complex interactions between upstream and downstream terrain features, thereby increasing the model's ability to account for local variations and irregularities in peak flow depth. Further, this approach could enable the model to generalize better across different terrain conditions by incorporating information from a larger spatial scale into the prediction process. 

To evaluate the effect of incorporating benchmark simulations, we compare the model’s performance with and without the additional benchmark simulation channel in section \ref{Sec: effects_of_benchmarkSimulation}.

\subsection{Data Generation}
We generated training and testing data using Titan2D at the Montecito site. Simulation results are dependent on a rainfall hyetograph, or time series of rainfall intensity, as well as a number of parameters that influence the initiation, growth, and characteristics of the debris flow. The simulations were performed for the entire study area by exploring various parameter combinations and recording the peak flow depth at each pixel as the target variable. We used elevation data from a post-event 1 m resolution digital elevation model derived from airborne lidar. The rainfall hyetograph was defined based on the observed rainfall that generated the debris flows on January 9, 2018. We used a 1-minute rainfall intensity time series derived from data recorded by the KTYD rain gauge, which is located roughly $5$ km west of the San Ysidro Creek watershed. We used the observed rainfall time series since it is typical of the short-duration, high-intensity bursts of rainfall that trigger postfire debris flows in this region \citep{kean2011situ}.

We fixed many of the model parameters based on values used in past studies \citep{tang2019evolution,mcguire2017debris}, but allowed six to vary. We identified these six parameters, namely \textit{(saturated hydraulic conductivity, Manning's roughness coefficient, the ratio of pore fluid pressure to total basal normal stress, the fraction of stream power effective in sediment detachment, effective grain size, and a rainfall intensity coefficient)}, based on factors known to have a strong influence on the initiation and growth of runoff-generated debris flows. Saturated hydraulic conductivity influences the infiltration capacity of the soil, with a greater value leading toless runoff. Manning's roughness coefficient influences flow resistance, particularly when sediment concentration is low. A greater value is associated with increased flow resistance. Sediment entrainment by overland flow is an increasing function of stream power. The fraction of stream power effective in sediment detachment therefore influences sediment detachment rates, particularly in areas of concentrated flow (e.g., channels). Here, the grain size distribution at our study site is characterized by a single effective grain size. This effective grain size influences sediment deposition rates by controlling the particle settling velocity. Lastly, we defined a rainfall intensity coefficient. This coefficient is multiplied by the observed rainfall intensity time series during the January, 9, 2018 debris flow event to generate rainfall time series that have different peak intensities. Increases in rainfall intensity promote increases in runoff, sediment entrainment, and debris flow volume \citep{mcguire2021time, gartner2014empirical}. All other model parameters were fixed. 

To train the model, we generated parameter combinations for Titan2D simulations that cover the parameter space (Table \ref{tab:parameter-range}).
However, randomly selecting parameter settings would have required an impractical number of simulations to cover the parameter space. Therefore we used a Latin Hypercube Sampling (LHS) method to choose parameter sets for Titan2D simulations \citep{mckay-lhs}. We applied the LHS method to generate 512 design points with different parameter combinations.

\begin{table}[h]
\caption{Input Parameter Range}
\label{tab:parameter-range}
\begin{tabular}{| m{7.5cm} | m{2cm} | m{2cm} |}
\hline
\textbf{Parameters }                                      & \textbf{Min Value} & \textbf{Max Value} \\ \hline
Saturated Hydraulic Conductivity                          & 1e-6      & 6e-6      \\ \hline
Manning Coefficient                                       & 0.03      & 0.1       \\ \hline
Ratio of pore fluid pressure to total basal normal stress & 0.5       & 0.8       \\ \hline
Fraction of stream power effective in sediment detachment & 0.005     & 0.05      \\ \hline
Effective Grain Size                                      & 5e-5      & 5e-4      \\ \hline
Rainfall Intensity Coefficient                            & 0.4       & 1.5      \\ \hline
\end{tabular}
\end{table}

We ran simulations on the Tufts HPC cluster and allocated 32 CPU nodes and 64 GB of memory for each simulation. The output of each simulation is a 2D array indicating peak flow depth for every pixel in the model domain raster. We further divided these arrays into 49 patches of size 256 x 256. This process yielded a dataset consisting of 25008 (512 x 49) images, which can be referenced by a combination of the patch location and parameter design point. 

For model development and evaluation, we partitioned this dataset into training and testing sets using a random sampling approach with an 80:20 split ratio. This resulted in a training dataset of 20,070 images and a testing dataset of 4,938 images. This division ensures a robust assessment of model performance while maintaining a balanced representation of the data.

\subsection{Data Pre-processing}\label{Sec: Preprocessing}
To ensure the compatibility and optimal performance of the simulation data within the model, it is essential to pre-process the data. This pre-processing involves two key steps: feature scaling and concatenating the input layer.

\subsubsection{Feature Scaling}
Feature scaling plays a crucial role in achieving better convergence during the training process and improved model performance. % It involves scaling various parameters to a consistent and appropriate range, which aids in achieving better convergence during the training process and improved model performance. 
We scaled the six Titan2D parameters that varied among simulations, the peak flow depth output from Titan2D, and the elevation data with the corresponding features of the benchmark simulation.
%
%As mentioned in the previous section, we allowed six Titan2d parameters to vary in our study. We scaled these six simulation parameters since they are of different orders of magnitude. 
%For better performance, we needed to scale and normalize the six Titan2d parameters  with the corresponding features of the benchmark simulation .
    \begin{equation}
        \hat{f}_{i,p}=\frac{f_{i,p} - f_{i,BM}}{f_{i,BM}}
    \end{equation}
    where $f_{i,p}$ is the $i^{th}$ parameter for the desired parameter combination $p$ for which we want to predict peak flow depth, and $f_{i,BM}$ is the corresponding $i^{th}$ parameter of the parameter combination used in the benchmark simulation. 
    
We scaled the peak flow depth output from Titan2d since it can range from $10^{-4}$ to 10 meters. %For our prediction to be accurate, we want to equally capture the entire range of flow. 
To effectively capture such a large range, we use a log scale to scale the peak flow depth. Specifically,
    \begin{equation}
        \hat{h}_{x,y} = log_{10}\left ( h_{x,y} \right )
    \end{equation}
    where $h_{x,y}$ is the maximum flow height at the grid-point $(x,y)$.
    
Lastly, the elevation can vary by orders of magnitude (e.g., $10^0-10^3$ m). For the Montecito site, elevation ranged from roughly $0 m $ to $1120 m$. While min-max scaling is commonly used for scaling elevations, we refrain from using it to maintain the transferability of our model across different terrains. Instead, we utilize a log-scale transformation to scale the elevation. Specifically, we let
    \begin{equation}
        \hat{z}_{x,y} = log_{10}\left ( z_{x,y} \right )
    \end{equation}
    where $z_{x,y}$ is the elevation at the grid-point $(x,y)$.

\subsubsection{Input layer Concatenation} 
In our model, we provide two distinct sets of inputs. The first is a 3-D array containing 2 channels of size $256 \times 256$. These channels are the peak flow depth of the patch simulated with the benchmark simulation parameters and the corresponding elevation of that patch, both of size $256 \times 256$. Both channels are scaled before concatenating into the input channel. This input channel serves as the input to the encoder part of the network.

For the second set, we reshape the parameter vector into a 1-D ordered array of length six. It is scaled as described in the previous section and fed into a fully connected dense network.

\subsection{Model Training}

The proposed model was trained on a dataset comprising 20,070 training samples, with the remaining 4,938 samples reserved for testing. The mean squared error (MSE) was adopted as the loss function due to its effectiveness in regression tasks. Specifically, the MSE was computed between the predicted peak flow depths from the model and those generated by the Titan2D simulations.

To optimize the training process, we utilized the Adam optimizer because of its demonstrated stability and efficiency across complex network architectures, including U-Net \citep{kingma2014adam}. The learning rate was tuned by evaluating a range of values ($1 \times 10^{-4}$ to $1 \times 10^{-6}$) using a grid search, with $1 \times 10^{-5}$ identified as the optimal value, balancing stable convergence and effective learning. The batch size was also tuned, with a value of 32 providing the best trade-off between computational efficiency and gradient stability. Training was conducted over 500 epochs, during which early stopping was applied to improve efficiency and prevent overfitting. Model performance was periodically evaluated on the testing dataset throughout the training phase. This systematic approach ensured that the model achieved optimal performance while maintaining generalization capabilities.

\subsection{Post Processing}\label{Sec: Postprocessing}
The prediction of maximum flow depth is provided in the logarithmic scale  to facilitate visualization and further analysis, it is necessary to convert these logarithmic values into a standard unit of measurement. 
Furthermore, to predict the peak flow depth across the entire map at specific parameter settings, we sequentially predict the maximum flow depths for each tiled patch of the map and then stitch them back together to reconstruct the prediction of the entire map.

\section{Results and Discussions}
%\improvement{Compare results to GP? This may be tricky but is a natural question.}
\subsection{Effects of Benchmark Simulations}\label{Sec: effects_of_benchmarkSimulation}

\begin{figure}[!h]
    \centering
    \includegraphics[width=\textwidth]{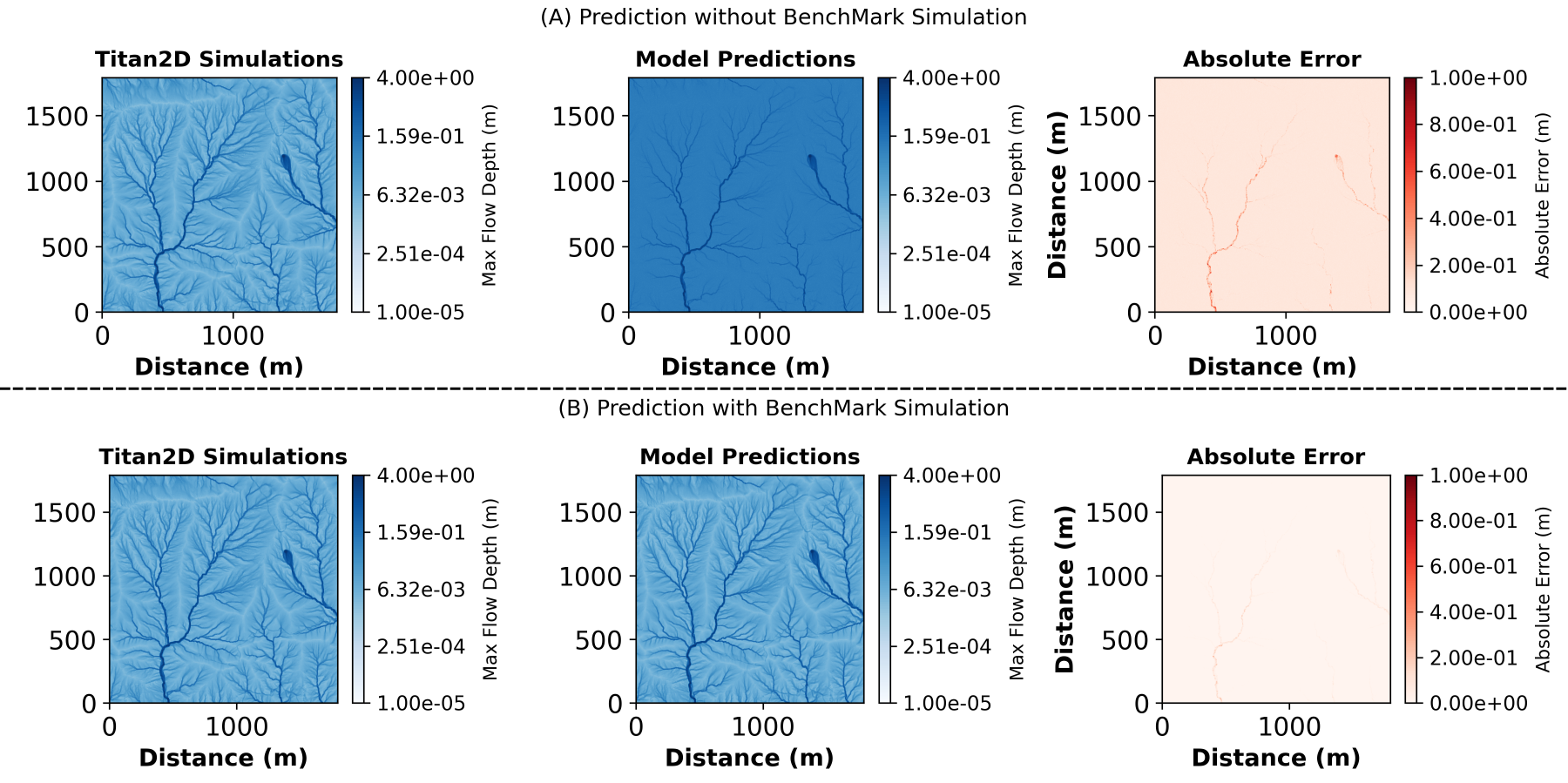}
    \caption{Comparison of model predictions with and without benchmark simulations. (A) Predictions without incorporating a benchmark simulation. (B) Predictions with a benchmark simulation included as an additional input channel, improving model performance. The absolute error plots highlight a reduction in errors when benchmark simulations are used, demonstrating the increased model performance from the upstream context.}
    \label{fig:benchmark_prediction_comparison}
\end{figure}

To assess the effects of incorporating benchmark simulations, we compare the model’s performance with and without the additional benchmark simulation channel. Visually, we observe that the model trained with benchmark simulations captures flow patterns more accurately, particularly in regions influenced by upstream conditions (Figure \ref{fig:benchmark_prediction_comparison}).

Error plots of Fig. \ref{fig:benchmark_prediction_comparison} illustrate the spatial distribution of prediction errors. The model without benchmark a simulation exhibits higher errors in areas where upstream flow dynamics strongly influence peak flow depth as compared to the prediction from a model using benchmark simulations. 

These results highlight the effectiveness of adding a benchmark simulation, which increases predictive accuracy by providing additional upstream flow context. This approach allows the model to better account for complex terrain interactions, ultimately improving its ability to generalize across diverse terrains.

\subsection{Emulator Performance and Validation}
We evaluated model performance by comparing peak flow depths, for a given parameter set, from a Titan2D simulation with the corresponding prediction from our model. Error maps are calculated as the absolute pointwise difference between the Titan2D results and the model predictions. 

\subsubsection{Error at the patch scale}
\begin{figure}[!h]
    \centering
    \includegraphics[width=\textwidth]{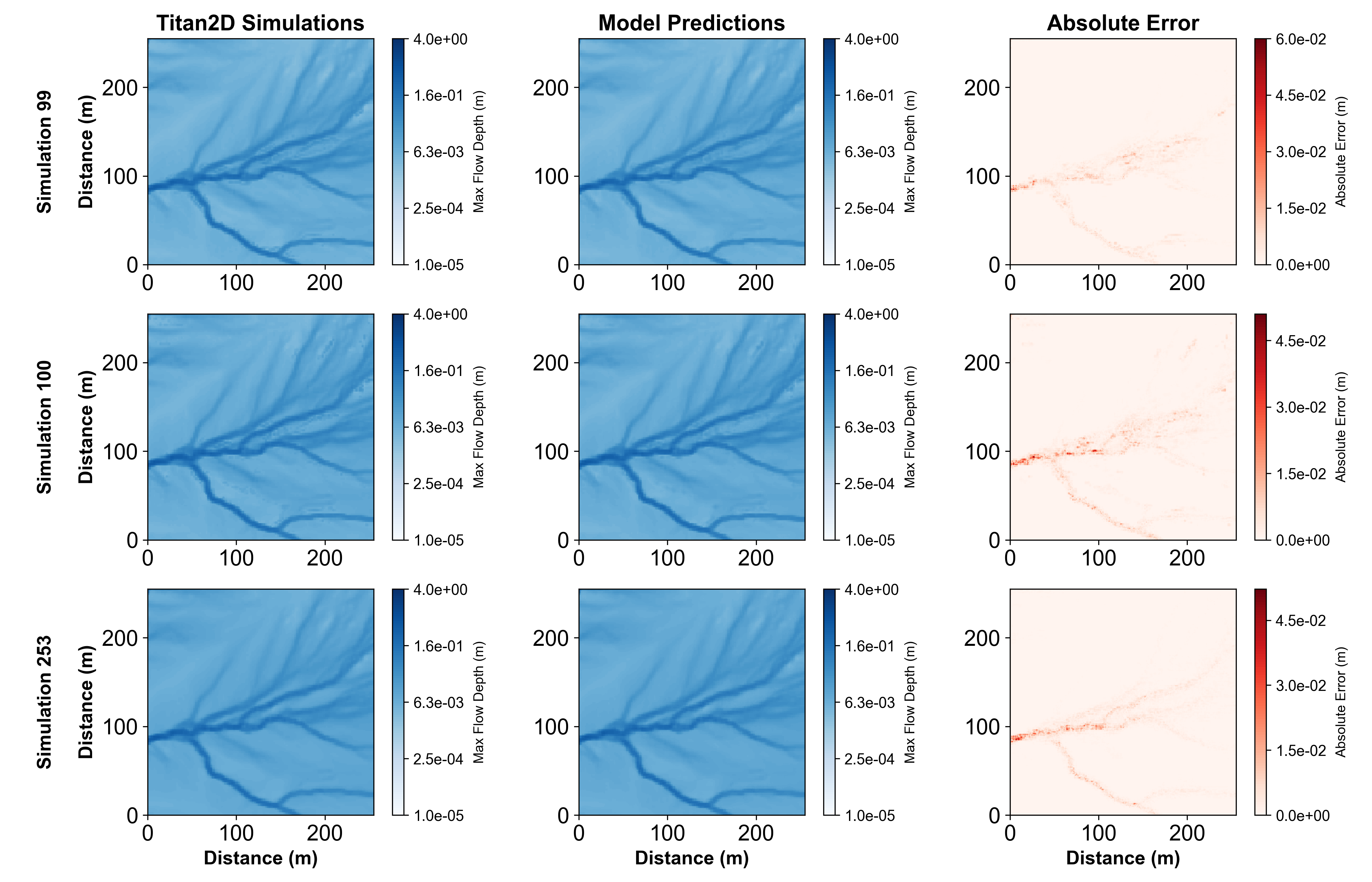}
    \caption{Comparison of patch data and model predictions across three arbitrary Titan2D simulations. Each subplot depicts patches extracted from Titan2D simulations alongside corresponding model predictions. Error is defined  at each point as the absolute value of the difference between the Titan2d simulations and the model prediction.}
    \label{fig:patch_comparison}
\end{figure}

As mentioned earlier, The model takes the concatenated input array of elevation and peak flow depth from the benchmark simulation for a given patch and predicts peak flow depth for the desired parameter set for the corresponding patch.
To validate the performance of our model, we tested the model on three unseen parameter sets not included in the training phase. The results demonstrate the model's ability to replicate the spatial patterns of peak flow depth observed in the Titan2D simulation and indicates that the model successfully generalizes to new parameter sets (Fig. \ref{fig:patch_comparison}). Maximum pointwise absolute error is below 10\%. This validation highlights the model's robustness and suitability for geophysical flow prediction tasks.

\subsubsection{Model Validation on Entire Study Area}
Building upon the validation performed on individual patches, we extended our assessment to the entire study area. The model's ability to predict peak flow depth across the full spatial domain was tested using a randomly generated parameter set, following the patch-predict-stitch approach detailed in Section \ref{Sec: Postprocessing}. In this approach, the study area is divided into smaller patches, predictions are made for each patch, and the results are stitched together to reconstruct the complete spatial prediction.

\begin{figure}[!hb]
    \centering
    \includegraphics[width=\textwidth]{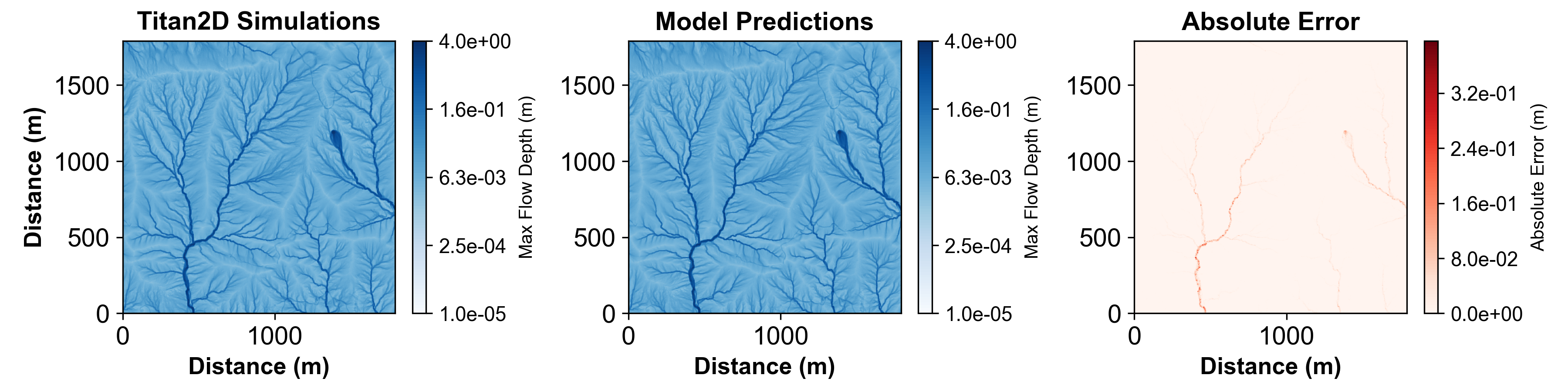}
    \caption{Comparison between the Titan2D simulation and model prediction over the entire study. Error refers to absolute error at each point.}
    \label{fig:map_comparison}
\end{figure}

The stitched output was compared to the Titan2D simulation results for the same parameter set to evaluate the accuracy of the predictions,. The model effectively captures the spatial variations in peak flow depth (Fig. \ref{fig:map_comparison}). The most significant errors occur in areas of concentrated flow, such as channels, where the flow depth is greatest. Even in these challenging regions, the magnitude of error is relatively small. Typical errors are on the order of 0.1 m compared to modeled peak flow depths of approximately 1 m. These errors are generally smaller than those reported for a Gaussian process model trained on Titan2d flow simulations across a model domain that included our Montecito study area as well as the downstream alluvial fan \citep{Spiller_McGuire_Patel_Patra_Pitman_2024}. Estimated errors in peak flow depth for the Gaussian process model were 0.16 m for a location on the alluvial fan where flow was not confined and 0.36 m in a channel with confined flow \citep{Spiller_McGuire_Patel_Patra_Pitman_2024}. The agreement between our model and Titan2d simulations underscores the model's potential as an accurate and efficient emulator for predicting flow dynamics over large spatial domains, making it a promising option for accelerating Monte Carlo calculations for real-world applications.

\subsubsection{Evaluating Flow Depth Predictions Across the Landscape}

\begin{figure}[!th]
    \centering
    \includegraphics[width=\textwidth]{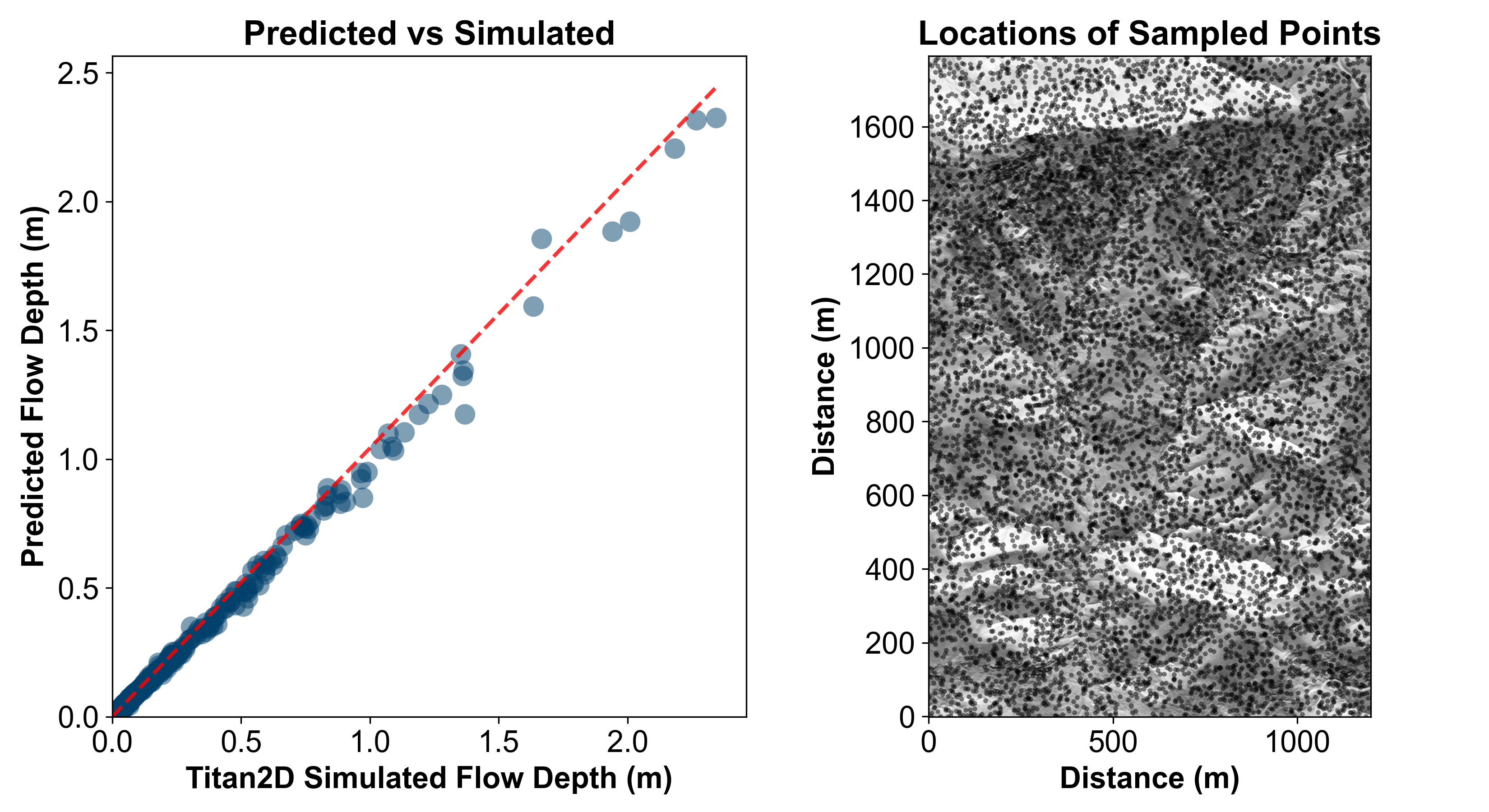}
    \caption{(Left) Scatter plot comparing the target flow height values obtained from Titan2D simulations (x-axis) with the predicted values from the deep learning surrogate model (y-axis). The red dashed line represents the 45° reference line, indicating perfect agreement between predictions and targets. (Right) Spatial distribution of the 20,000 randomly sampled points used for the scatter plot analysis overlayed on the hillshade visualization of the Montecito site. }
    \label{fig:scatter_plot}
\end{figure}

To further evaluate the accuracy of our deep learning surrogate model, we compared its output to the corresponding peak flow depth from Titan2D simulations. A total of 20,000 points were randomly sampled across the study area from over 10 independent simulations, each performed using previously unseen parameter combinations (Fig. \ref{fig:scatter_plot}). This sampling approach ensured that the validation process included a diverse set of flow conditions and spatial locations, thereby testing the generalizability of the surrogate model beyond its training data. Results demonstrate a strong agreement between the model and Titan2d simulations (Fig. \ref{fig:scatter_plot}). 

Additionally, the Pearson-R coefficient value between the model’s predictions and the Titan2D simulations was approximately 0.9988, indicating an almost linear correlation between the predicted and simulated peak flow depths. This alignment indicates the model's capability to accurately replicate the complex flow dynamics simulated by Titan2D. 

Across all validation studies, the surrogate model demonstrates strong agreement with Titan2D simulations, suggesting its suitability for tasks requiring rapid and accurate predictions. By substantially reducing computational costs compared to physics-based simulations, the surrogate model approach enables the exploration of a broader range of parameter combinations and uncertainty scenarios. We note that we are using the data driven U-net model as a surrogate of the full physics-based solver and not a replacement thereof as in physics-informed neural networks. 

\subsection{Uncertainty Quantification}

\begin{figure}[!t]
    % \begin{subfigure}{0.49\textwidth}
    % \centering
    %     \includegraphics[width=\textwidth]{figures/Figure3_ensemble_mean.png}
    %     \caption{Ensemble Mean}
    %     \label{fig:mean_UQ}
    %     \end{subfigure}
    % % \hspace{0.08\textwidth}
    % \begin{subfigure}{0.49\textwidth}
    % \centering
    %     \includegraphics[width=\textwidth]{figures/Figure4_ensemble_standard_deviation.png}
    %     \caption{Ensemble Standard Deviation}
    %     \label{fig:std_UQ}
    % \end{subfigure}
    \centering
    \includegraphics[width=\textwidth]{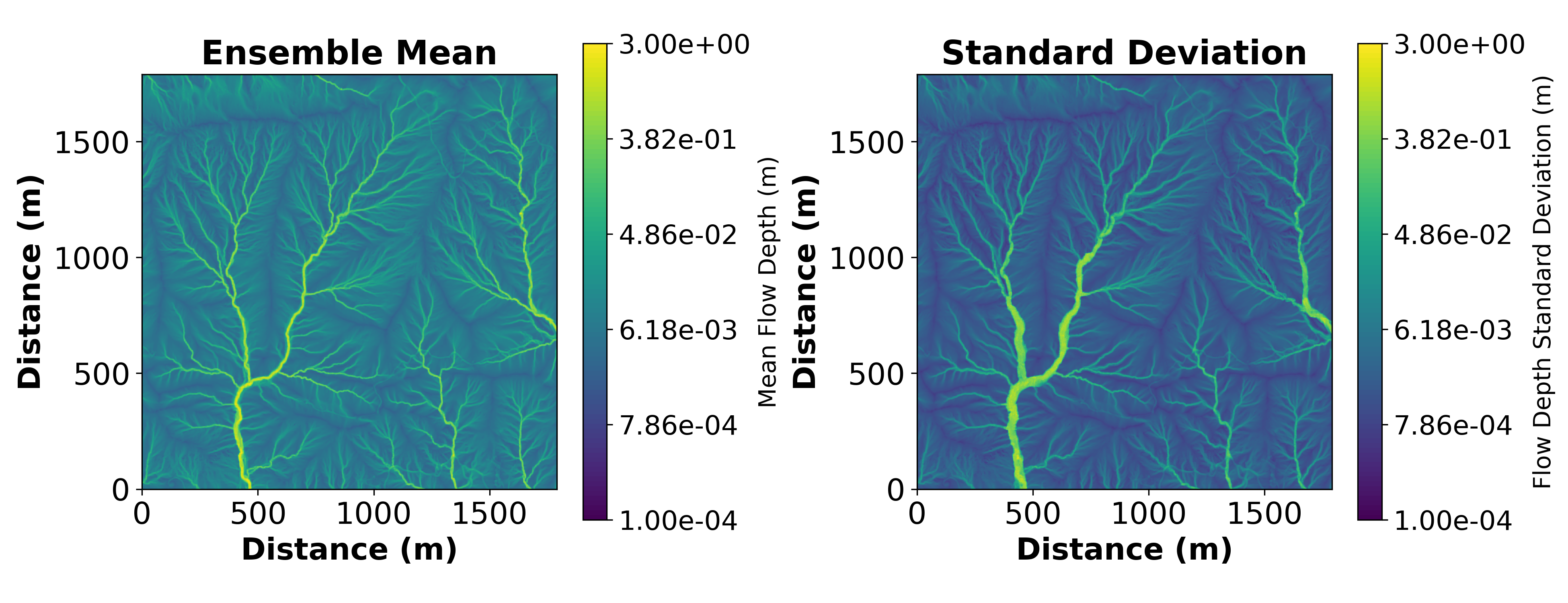}
    
    \caption{Spatial uncertainty quantification of the predicted peak flow depth across the entire Montecito study area based on an ensemble of 10,000 Monte Carlo simulations. (a) The ensemble mean represents the average predicted peak flow depth.(b) The standard deviation indicates the variability in peak flow depth predictions at each location.}
    \label{fig:UQ}
\end{figure}

%Understanding and quantifying the uncertainties associated with model outputs are essential for making informed decisions and drawing reliable conclusions. 
A computationally efficient emulator enables the use of large ensembles for uncertainty calculations by rapidly evaluating the model across a large parameter space. We illustrate  by using a Monte Carlo  approach to quantify effect of variability and uncertainty in the input parameters (Table \ref{tab:parameter-range})  in  peak flow depth predictions. 

%The analysis focused on predicting the maximum flow depth across the study area. 
The Monte Carlo simulation approach randomly selects 10,000 input parameter sets from the ranges specified in Table \ref{tab:parameter-range}. The patch-predict-stitch methodology (described in Section \ref{Sec: Postprocessing})  generates an ensemble of spatial predictions. To evaluate the uncertainty, the ensemble predictions were analyzed by calculating the mean and standard deviation of the predicted peak flow depths at each spatial location (Fig. \ref{fig:UQ}). 

The ensemble mean (Fig. \ref{fig:UQ} left) provides an average representation of the predicted flow depths, highlighting the dominant flow patterns under the given parameter ranges. It provides an estimate for typical peak flow depths and how they could vary in space across the study area. The standard deviation (Fig. \ref{fig:UQ} right) can be used to highlight areas with greater uncertainty in peak flow depth. Areas with a higher standard deviation, which often correspond to channels or regions with steep gradients in flow depth (Fig. \ref{fig:UQ}), are locations where uncertainty in the input parameters could lead to substantial variability in predicted peak flow depth. %Ensemble mean and standard deviation provide basic estimates for typical outcomes and uncertainties, associated with variations in input parameters, across the study area. These metrics could aid in identifying regions where predictions are particularly sensitive to variability in input parameters. 
Such analyses are crucial for increasing the robustness of the model as an emulator and ensuring its applicability in real-world hazard assessments and decision-making processes.

\subsection{Threshold Exceedance Probabilities}
Probabilistic analyses can be an important component of debris flow hazard and risk assessments \citep{barnhart2023user,oakley2023toward}. Unlike deterministic approaches that provide a single prediction of peak flow depth, probabilistic maps offer a more comprehensive understanding of uncertainty and risk associated with hazardous events by illustrating the likelihood of flow depth exceeding critical thresholds. This information is especially valuable for understanding and managing risks in regions prone to debris flows.

Building upon the uncertainty quantification analysis conducted in the previous section, we produced a map showing the probability of the peak flow depth exceeding a predefined threshold. For illustration of methodology, we select a threshold of $0.3$ m as a critical flow depth that, when exceeded, could represent an increase in hazard.

Utilizing the ensemble generated from the previous section, we calculated the probability of the peak flow depth surpassing the threshold at each point on the map as

\begin{equation}
    P_{i,j} = \frac{\sum_{n=1}^{N}1_{\hat{h}_{i,j,n} \geq h_{\theta}}}{N}
\end{equation}

\begin{figure}[!h]
    \centering
    \includegraphics[width=0.9\textwidth]{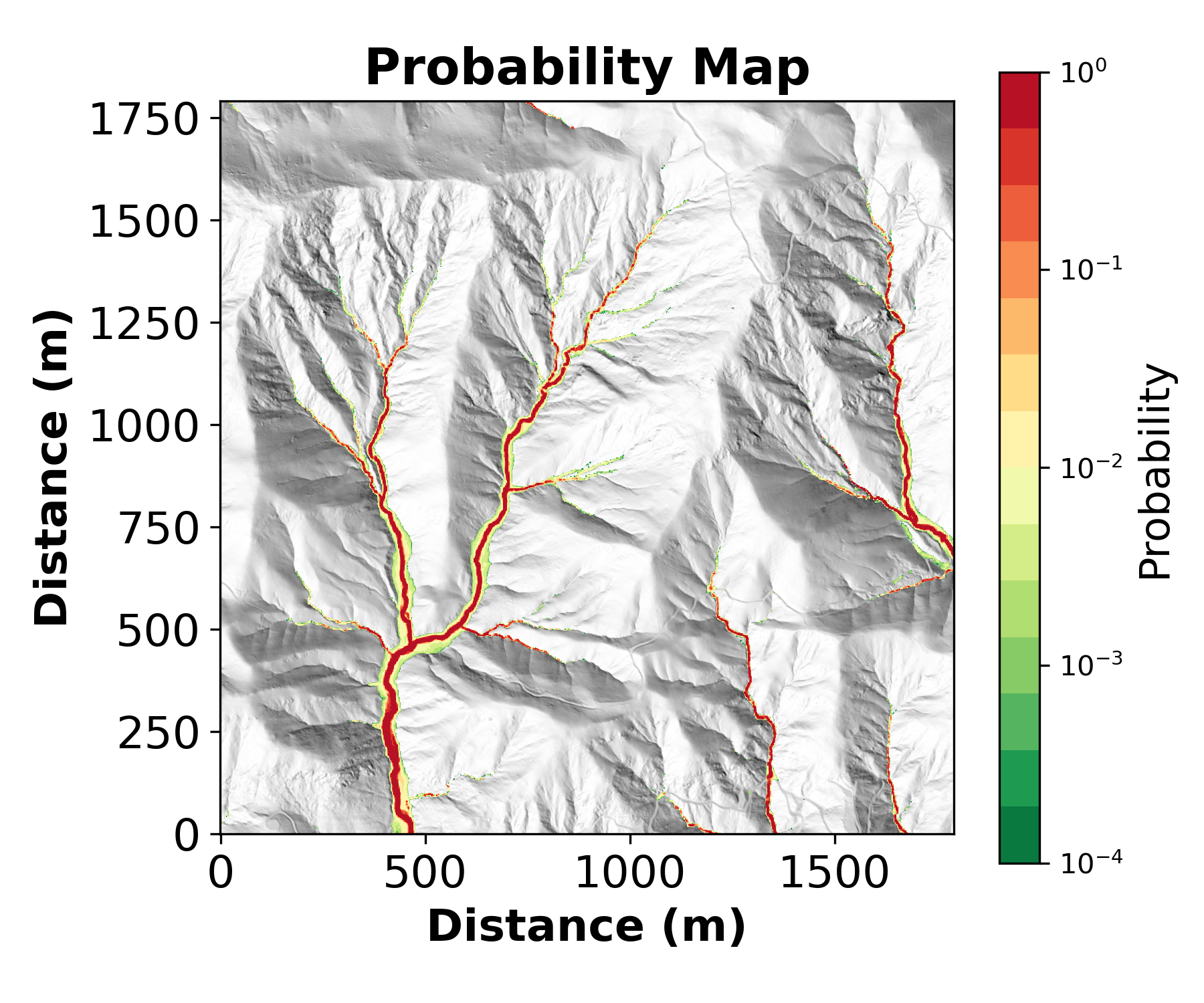}
    \caption{Probability map illustrating the likelihood of peak flow depths exceeding a threshold of 0.3 m across the study area ($P(h>h_\theta=0.3)$. This map was generated using an ensemble of 10,000 Monte Carlo simulations.}% providing a highlight of areas with heightened probability of flow exceeding a threshold $P(h > h_\theta)$.}
    \label{fig:ProbMap_thomas}
\end{figure}

In the above equation, $P_{i,j}$ represents the probability of the peak flow depth exceeding the threshold $h_{\theta}$ at point $\left(i,j\right)$, $N$ is the total number of ensembles, $\hat{h}_{i,j,n}$ denotes the maximum flow depth for point $\left(i,j\right)$ for the $n^{th}$ ensemble, and $1_{\hat{h}_{i,j,n}}$ is the indicator function that equals 1 if $\hat{h}_{i,j,n}$ is greater than the threshold $h_{\theta}$, and 0 otherwise. Applying this equation across the entire study area, we generated a map the likelihood of encountering peak flow depths beyond the specified threshold (Fig. \ref{fig:ProbMap_thomas}). This map highlights areas with varying probabilities of surpassing the critical flow depth threshold. This and similar probabilistic output from ensemble forecasts, which are made possible by the computational efficiency of the emulator relative to the Titan2d simulations, could provide insight into spatial variations in hazard intensity. Exceeding critical values of flow depth or velocity, for example, have been used as metrics to assess hazard intensity \citep{hurlimann2008evaluation,raetzo2002hazard}. This information can support decision-makers in prioritizing mitigation strategies, allocating resources, and improving disaster preparedness plans for communities at risk.

\subsection{Case Study: Oak Creek site}

\begin{figure}[!hb]
    \centering
    \includegraphics[width=\textwidth]{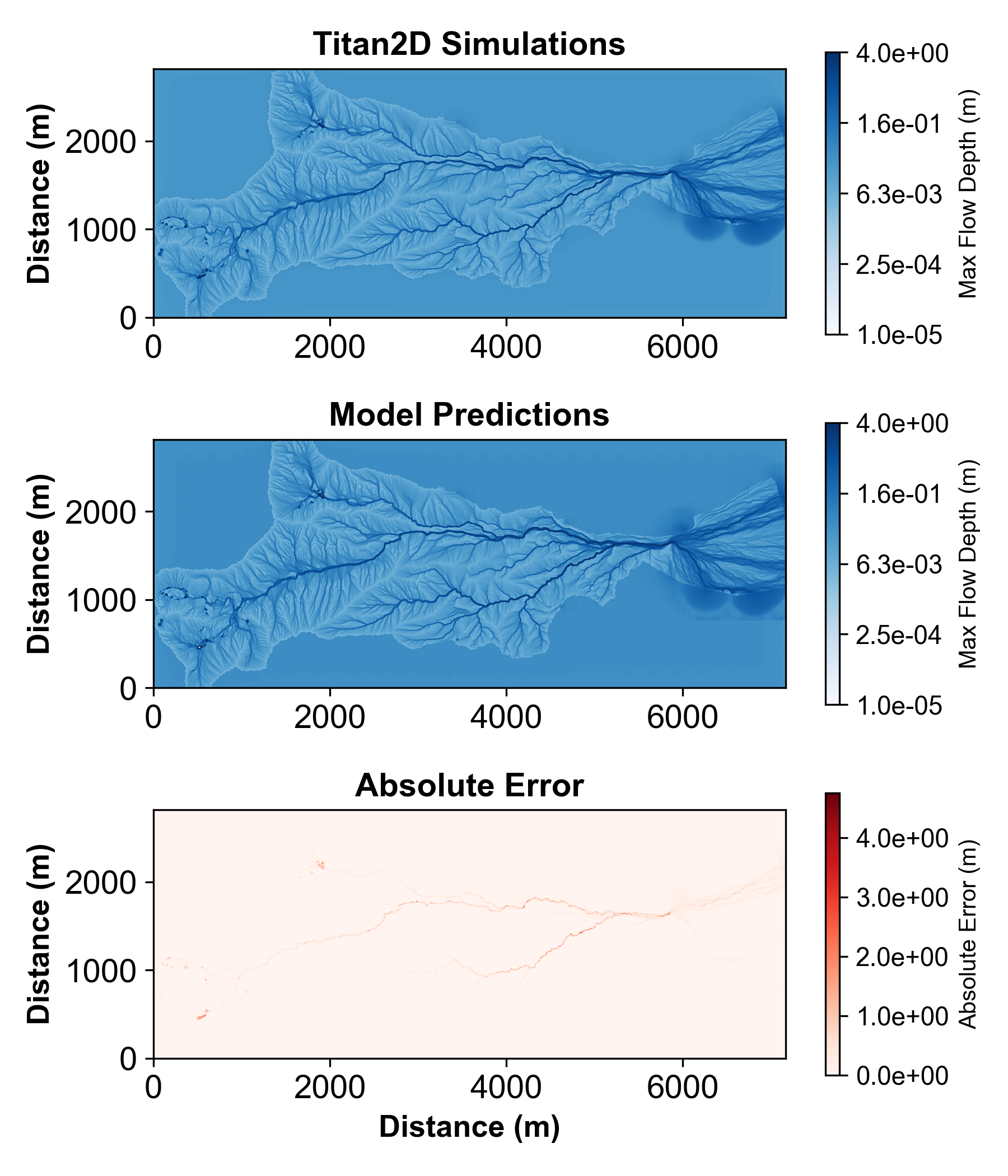}
    \caption{Comparison of Titan2D simulation and model prediction from our deep learning-based surrogate model for peak flow depth across the complex terrain of the Oak Creek site.}
    \label{fig:Arizona_Predictions}
\end{figure}

To further assess the performance our surrogate modeling approach across landscapes with different characteristics, we conducted a case study at the Oak Creek site. This site is characterized by more complex terrain relative to the Montecito site and presents several challenges. In particular, the model domain is larger and there is a greater diversity of terrain features. The model domain includes a combination of steep hillslopes, confined channels, and a gentler alluvial fan with varying degrees of flow confinement. We expect these different terrain features to lead to different flow dynamics. In contrast, simulations at the Montecito site focused on headwater drainages with steep hillslopes and channels where the flow was relatively confined.

Data for the Oak Creek site were generated using Titan2D simulations, employing the same parameter ranges outlined in Table \ref{tab:parameter-range}. Preprocessing steps, including segmentation of the terrain into patches of size $256 \times 256$ pixels, followed the preprocessing methodology described in Section \ref{Sec: Preprocessing}. The model utilized the same modified U-Net architecture and training methodology as in previous experiments of the Montecito site. Model predictions of peak flow depths compare well with Titan2d simulations across different parts of the landscape, including steep hillslopes, confined channels, and the flatter alluvial fan (Fig. \ref{fig:Arizona_Predictions}).

\begin{figure}[!hb]
    \centering
    \includegraphics[width=\textwidth]{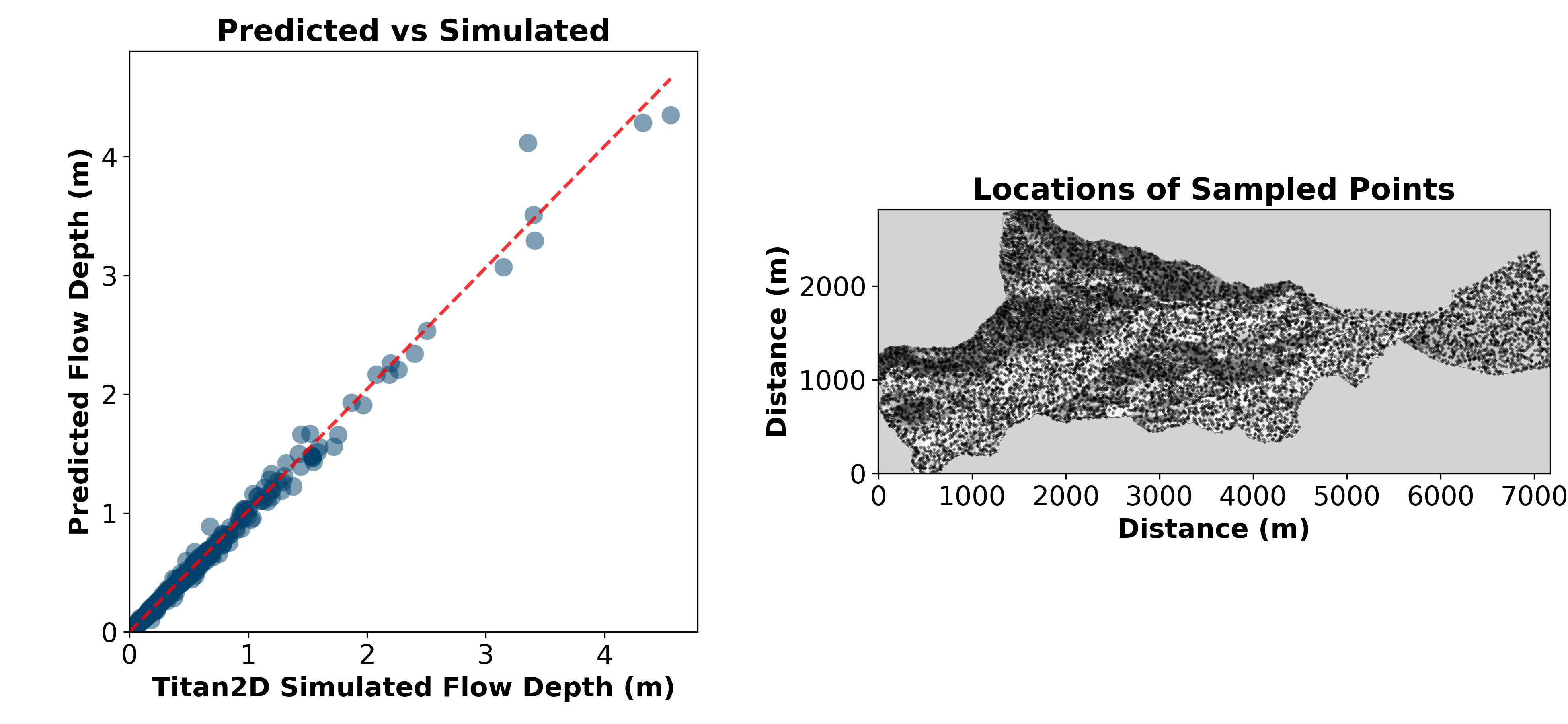}
    \caption{Scatter plot comparing the target flow height values obtained from Titan2D simulations (x-axis) with the predicted values from the deep learning surrogate model (y-axis) on Oak Creek site. The red dashed line represents the 45° reference line, indicating perfect agreement between predictions and targets. (Right) Spatial distribution of the 20,000 randomly sampled points used for the scatter plot analysis overlayed on the hillshade visualization of the terrain.}
    \label{fig:arizona_scatterplot}
\end{figure}

Similar to the evaluation performed at the Montecito site, we further evaluated the accuracy of our deep learning surrogate model by comparing its output to the corresponding peak flow depth from Titan2D simulations. A total of 20,000 points were randomly sampled across the study area from over 10 independent simulations, each performed using previously unseen parameter combinations. Results demonstrate a strong agreement between the model and Titan2d simulations with Pearson-R coefficient of 0.9975 (Fig. \ref{fig:arizona_scatterplot}). This alignment further endorses the model's capability to accurately replicate the flow dynamics simulated by Titan2D on more complex terrain.

Performing a case study on a new site, such as Oak Creek, with terrain characteristics that differ from the site where the model was initially developed is beneficial for evaluating the generalizability of the model. Results from the case study demonstrate that our surrogate model is not performing well only at the specific site where it was developed, but it is capable of providing accurate predictions across a broader range of geographic settings. Moreover, testing on more complex and varied terrain, such as that from Oak Creek, also serves to provide confidence for real-world hazard and risk assessment applications.

\section{Conclusions}
In this study, we presented a novel approach utilizing deep learning-based surrogate modeling for the emulation of debris flow simulations. The proposed framework leverages a modified U-Net architecture with necessary modifications to account for the nature of the flow data to predict peak flow depths across diverse terrains, significantly reducing computational costs compared to conventional physics-based simulation models, such as Titan2D.
We demonstrated the effectiveness and reliability of our model in accurately predicting peak flow depths across diverse terrain. Our approach was tested on two distinct sites, which we refer to as Montecito and Oak Creek, featuring varying terrain conditions across steep headwater drainages and alluvial fans.
The results from the Montecito site demonstrate the model's ability to accurately replicate flow dynamics in steep headwater drainages where debris flows initiate and grow. Building on this, the case study at the Oak Creek site validated the model's robustness and generalizability by successfully capturing flow patterns across a domain with more complex and diverse terrain features, including an alluvial fan. The model consistently exhibited high accuracy in predicting peak flow depths and spatial flow patterns, as evidenced by the close agreement between model predictions and Titan2D simulations across both sites. 

In addition to demonstrating predictive accuracy, our framework employs the patch-predict-stitch methodology where the study area was divided into smaller patches for prediction. Patches were subsequently combined to reconstruct spatially continuous results over the entire domain. We note our careful use of a single global flow dataset, which we refer to as a benchmark simulation, to provide a mechanism for improving the ``patch-predict-stitch" method. This approach highlights its scalability for large terrain maps, which is needed for hazard and risk assessments at regional scales.

Furthermore, the significant reduction in computational costs achieved by the surrogate model compared to traditional physics-based simulations, enables the exploration of a wide range of parameter combinations and scenarios, paving the way for more efficient and informed hazard analyses. Uncertainty quantification using Monte Carlo simulations provided critical insights into spatial variability and sensitivity to input parameters. The ensemble analysis revealed spatial patterns of uncertainty, highlighting regions with higher prediction variability. We further demonstrated how the surrogate model could be employed to assess the likelihood of critical flow thresholds being exceeded, which could be particular beneficial for assessing hazard intensity. 

In conclusion, this deep learning-based surrogate modeling has the potential to be a powerful tool for emulating geophysical flows. Here, we focus on debris flow simulations but expect the methodology could generalize to simulating other types of geophysical flows. Our investigation demonstrated that a modified U-Net architecture can effectively capture complex relationships between input parameters, terrain characteristics, and flow dynamics. This study highlights the promise of deep learning-based surrogate models as accurate, efficient, and reliable tools for geophysical flow prediction and hazard assessment. Future work could explore the integration of transfer learning techniques to further improve model performance on terrain with unique or unseen characteristics. Moreover, training with models other than Titan2D would allow us to capture flows with a wider range of physics in our surrogate model. Overall, the proposed approach represents a step forward in addressing challenges associated with modeling geophysical flows for hazard and risk assessments by bridging the gap between computational efficiency and prediction accuracy.

\section{Code Availability}
The code for building, training, and evaluating our deep learning model is available on \url{https://gitlab.com/palakkpatel/nn_prob_map}.

The source code for Titan2D, used for generating simulations, is available on \url{https://github.com/TITAN2D/titan2d}.
\section{CRediT author statement}
Palak Patel: Conceptualization, Methodology, Software,   Writing- Original draft preparation.  Abani Patra: Supervision, Conceptualization, Methodology, Writing -- Reviewing and Editing, Validation.:   Luke Mcguire: Conceptualization, Writing- Reviewing and Editing, Data curation,Validation.
\section{Acknowledgments}
This research was funded by the National Science Foundation under Grant No. NSF-2004302. 

\bibliographystyle{cas-model2-names}
\bibliography{ref}
\appendix
\centerline{\bf Appendix}
\section{Flow Model Details}
The physics-based model for flow and sediment transport used here is a Titan2d implementation of the model presented in \citeauthor{mcguire2017debris} \citeyear{mcguire2017debris} \citep{Spiller_McGuire_Patel_Patra_Pitman_2024}. The model represents rainfall, infiltration, fluid flow, and sediment entrainment and deposition processes. {The model couples the processes of fluid flow, entrainment and deposition, and topographic change, such that the topography can evolve during a rainstorm in response to flow.} The governing equations are solved within the Titan2D framework \cite{titan2d2005, titan2d2019} which employs an adaptive mesh, finite volume scheme to solve hyperbolic PDEs describing shallow-water type mass flows over digital elevation models of real topography. 

The equations representing the motion of fluid and sediment can be written as a set of depth averaged conservation laws,
 
\begin{equation}
    \frac{\partial \mathbf{U}}{\partial t} + \frac{\partial \mathbf{F}}{\partial x} + \frac{\partial \mathbf{G}}{\partial y} = \mathbf{S_0} + \mathbf{S_1} + \mathbf{S_2},
\end{equation}
 
where 

$    \mathbf{U} = \{
    h,   uh,  vh, c_1h,  \cdots  c_kh
    \}^{T},
    \mathbf{F} = \{ 
    hu,   hu^2 + \frac{1}{2}g_zh^2,  huv,  huc_1,  \cdots  huc_k,
    \}^{T}, 
    \mathbf{G} = \}
    hv,   huv, hv^2 + \frac{1}{2}g_zh^2,  hvc_1,  \cdots  hvc_k,
    \}^{T},
$
and where $h$, $u$, $v$, and $c_i$ are flow depth, velocity along $x$-axis, velocity along $y$- axis, and sediment concentration of particle size class $i$. Components of gravitational acceleration in the $x$, $y$, and $z$ directions are given by $g_x$, $g_y$, and $g_z$, respectively, and $k$ denotes the number of particle size classes. $\mathbf{S_0}$, $\mathbf{S_1}$ and $\mathbf{S_2}$ are source terms. 
$\mathbf{S_0}$ denotes the contributions of mass sources and sinks associated with the effective rainfall rate, $P_{eff}$, and the soil infiltration capacity, $I$, as well as momentum sources and sinks arising from variations in topographic elevation, and spatial variations in sediment concentration and debris flow resistance terms, $S_x$ and $S_y$.
$\mathbf{S_1}$ accounts for flow resistance using a depth-dependent Manning’s formulation, and is given as    
$ 
    \mathbf{S_1} = \{
    0,
    g_z\eta ^2hu\sqrt{hu^2+hv^2}/h^{7/3},
    g_z\eta ^2hv\sqrt{hu^2+hv^2}/h^{7/3}, 
    0, \cdots,
    0
    \},
$
where {$\eta$} is the Manning friction coefficient. The friction coefficient varies with flow depth according to
 
\begin{equation}
    \eta=
    \begin{cases}
        \eta_0\left(h/h_c\right)^{-\epsilon} & h \le h_c\\
        \eta_0 & h > h_c
    \end{cases},
\end{equation}
 
where $\eta_0$ is the hydraulic roughness coefficient, $h_c$ is a critical flow depth and $\epsilon$ is a phenomenological exponent.
Soil infiltration capacity, $I$, is represented by the Green-Ampt model where
\begin{linenomath*}
\begin{equation}
    I=k_s\frac{Z_f+h_f+h}{Z_f},
\end{equation}
\end{linenomath*}
with $k_s$ denoting saturated hydraulic conductivity, $h_f$ the wetting front potential, $Z_f=V/(\theta_s-\theta_i)$ the depth of the wetting front, $V$ the cumulative infiltrated depth, $\theta_s$ the volumetric soil moisture content at saturation, and $\theta_i$ the initial volumetric soil moisture content.  
The source term $\mathbf{S_2}$ accounts for sediment entrainment and deposition processes \cite{hairsine1992modeling,hairsine1992rill}. 
{The topographic surface evolves in response to sediment entertainment and deposition according to} 
 
\begin{equation}
\frac{\partial z}{\partial t}=\frac{1}{\rho_s(1-\phi)}\left(\sum_{k=1}^K{d_k-e_k-e_{rk}-r_k-r_{rk}}\right).
\end{equation}

{Here, $\phi$ is the bed sediment porosity and $e_k$ and $e_{rk}$ denote the sediment detachment and redachment rates due to raindrop impact as defined by} \citeauthor{mcguire2016constraining} \citeyear{mcguire2016constraining}. 
\end{document}